\documentclass[10pt,twocolumn,superscriptaddress,amssymb,amsmath,aps,prl]{revtex4-2}
\usepackage{graphicx}

\usepackage{bbm}

\usepackage[normalem]{ulem}

\usepackage{color}

\newcommand{\cev}[1]{\reflectbox{\ensuremath{\vec{\reflectbox{\ensuremath{#1}}}}}}

\renewcommand{\tablename}{\textbf{Table}}

\renewcommand{\thetable}{\arabic{table}}
\makeatletter
\def\@caption@fignum@sep{ $|$ }
\def\fnum@figure{\textbf{Fig.}\nobreakspace\textbf{\thefigure}}
\def\fnum@table{\tablename\nobreakspace\textbf{\thetable}}
\makeatother

\begin{document}

\setcitestyle{super}

\title{Atomic Dirac energy-level dynamics and redshift in the $\textbf{4}\boldsymbol{\times}\textbf{U(1)}$ gravity gauge field}
\date{June 27, 2025}
\author{Mikko Partanen}
\affiliation{Photonics Group, Department of Electronics and Nanoengineering, 
Aalto University, P.O. Box 13500, 00076 Aalto, Finland}
\author{Jukka Tulkki}
\affiliation{Engineered Nanosystems Group, School of Science, Aalto University, 
P.O. Box 12200, 00076 Aalto, Finland}

\begin{abstract}
Gravitational interaction unavoidably influences atoms and their electromagnetic radiation field in strong gravitational fields. Theoretical description of such effects using the curved metric of general relativity is limited due to the classical nature of the metric and the assumption of the local inertial frame, where gravitational interaction is absent. Here we apply unified gravity extension of the Standard Model [Rep.~Prog.~Phys.~\textbf{88}, 057802 (2025)] to solve the Dirac equation for hydrogen-like atoms in the $4\times \mathrm{U(1)}$ gravity gauge field, which appears alongside all other quantum fields. We show that the gravity gauge field shifts the atomic Dirac energy levels by an amount that agrees with the experimentally observable gravitational redshift. Our result for the redshift follows directly from quantum field theory and is strictly independent of the metric-based explanation of general relativity. Furthermore, we present how gravitational potential gradient breaks the symmetry of the electric potential of the atomic nucleus, thus leading to splitting of otherwise degenerate spectral lines in strong gravitational fields. Enabling detailed spectral line analysis, our work opens novel possibilities for future investigations of quantum photonics phenomena in strong gravitational fields.
\end{abstract}

\maketitle



\begin{figure}
\centering
\includegraphics[width=0.98\columnwidth]{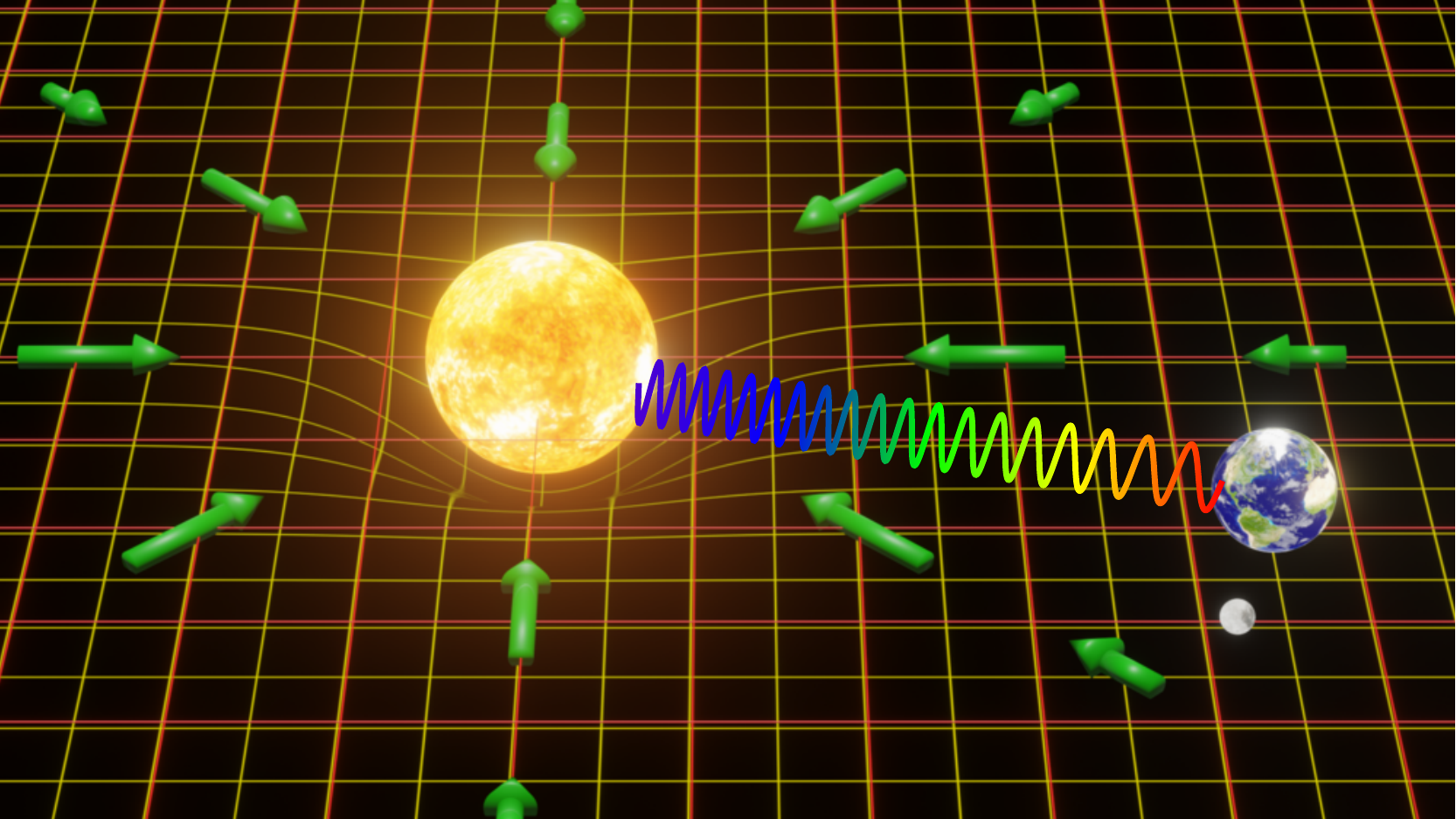}
\caption{\label{fig:illustration}
Illustration of the gravitational redshift of light escaping from the gravitational potential of a star. Light appears redshifted when detected at the higher gravitational potential of the Earth. We prove that the redshift of spectral lines is directly obtained by solving the Dirac equation of atoms in the $4\times \mathrm{U(1)}$ gravity gauge field.}
\end{figure}

One of the greatest triumphs of the equivalence principle of general relativity (GR) is the prediction that gravity influences the passage of time \cite{Bothwell2022,Roura2020,Misner1973,Moore2013,Einstein1916,Einstein1907,Einstein1911}. Consequently, radiation emitted by atoms appears redshifted when observed at a higher gravitational potential as illustrated in Fig.~\ref{fig:illustration}. After its first experimental confirmation by Pound and Rebka in 1959 \cite{Pound1959}, gravitational redshift has become crucial in astrophysics and cosmology, influencing the interpretation of spectra of white dwarfs \cite{Chandra2020,Joyce2018,Grould2017}, neutron stars \cite{Tang2020}, and objects near black holes \cite{Abuter2018,Zucker2006}. Gravitational redshift also contributes to the precision tests of GR and other theories of gravity \cite{Will2014,Zheng2023,Pumpo2021,Ufrecht2020,GonzalezHernandez2020,Herrmann2018,Delva2018,Wolf2016,Kraniotis2021}. The effects of gravity on atoms in GR have also been theoretically investigated \cite{Parker1980,Parker1982,Bringmann2023,Wanwieng2023,Kanno2025}. Numerous experiments have confirmed the predictions of GR, but only up to \emph{the first power of the gravitational constant} \cite{Will2014,Turyshev2007,Turyshev2009b}. This raises the question whether GR remains accurate beyond this limit.

In GR, gravity influences classical dynamics of massive bodies and light waves through the curved metric governed by the stress-energy-momentum tensor source term in the Einstein equation. However, GR remains a stand-alone theory isolated from the other fundamental interactions described by the quantum fields of the Standard Model \cite{Schwartz2014,Peskin2018}. The same applies to parametric modifications of GR \cite{Brans1961,Capozziello2007,Capozziello2010,deRham2011,Jimenez2018b,Soares2025}. The quantum field theory treatment of GR has only been developed as a low-energy effective field theory due to the inherent nonrenormalizability of GR \cite{Bambi2023,Casadio2022,Donoghue1994b,Donoghue1999,Donoghue2002,Stelle1977,Schwartz2014,Rocci2024}. Alternative approaches based on different ideas, such as string theory \cite{Becker2007,Dine2007,Green1987} and loop quantum gravity \cite{Ashtekar1986,Jacobson1988,Rovelli1990,Rovelli2008}, are under development.

In contrast, a recently introduced quantum field theory, unified gravity (UG) \cite{Partanen2025a}, describes gravity by the metric-independent $4\times \mathrm{U(1)}$ tensor gauge field, which appears as an extension of the Standard Model. In the semiclassical limit, it allows for dynamical description of all the same phenomena, which are calculated through the metric in GR. On the relation between UG and GR, we only mention that teleparallel equivalent of GR \cite{Bahamonde2023a,Aldrovandi2012,Maluf2013} results from one special geometric condition of UG \cite{Partanen2025a}. However, the pertinent geometric condition \emph{breaks the $4\times \mathrm{U(1)}$ gauge symmetry} and makes teleparallel equivalent of GR profoundly different from the Minkowski spacetime formulation of UG used in this work.

The coupled dynamical field equations of UG include only \emph{known physical constants}, they are written entirely within a global Minkowski frame, and the gravity gauge field enters coherently to all other physical fields of the Standard Model, contrasting the metric approach of GR. At an arbitrary point of spacetime in UG, one can place an atomic clock that measures the local time. In contrast to GR, UG does not assume the local inertial frame. The energy levels of the atomic clock can be calculated using the Dirac equation of UG. We show that the gravity gauge field shifts the energy levels of atoms by an amount that agrees with the experimentally observable gravitational redshift. Therefore, UG provides the quantum-field-theory explanation for gravitational redshift that is entirely different and independent of the metric-based explanation of GR. The first-order predictions of the two theories in powers of the gravitational constant are equal, but \emph{the theories differ regarding the higher-order corrections}, yet to be experimentally measured \cite{Will2014,Zheng2023,Pumpo2021,Ufrecht2020,GonzalezHernandez2020,Herrmann2018,Delva2018,Wolf2016,Kraniotis2021}. The other benchmark effects, the gravitational lensing and the perihelion precession of planetary orbits, are investigated using UG in preprints \cite{Partanen2025c,Partanen2025d}.

\section{Gauge field of unified gravity}
\vspace{-0.3cm}

We use a semiclassical approach in which the gravitational field is treated classically. Here we briefly review the solution of the UG gravity gauge field for a classical point mass in the Minkowski metric \cite{Partanen2025c}. We use the global Cartesian Minkowski frame coordinates $x^\mu=(ct,x,y,z)$, where $c$ is the speed of light in vacuum and in zero gravitational potential. The components of the diagonal Minkowski metric tensor $\eta_{\mu\nu}$, are given by $\eta_{00}=1$ and $\eta_{xx}=\eta_{yy}=\eta_{zz}=-1$. The Einstein summation convention is used for all repeated indices below.

In the global Minkowski frame, the field equation of gravity in the harmonic gauge of UG, $P^{\mu\nu,\rho\sigma}\partial_\rho H_{\mu\nu}=0$, is given by \cite{Partanen2025a,Partanen2025c}
\begin{equation}
 -P^{\mu\nu,\rho\sigma}\partial^2H_{\rho\sigma}=\kappa T_\mathrm{m}^{\mu\nu}.
 \label{eq:UGgravity}
\end{equation}
Here $H_{\rho\sigma}$ is the gravity gauge field, $\partial^2=\partial^\rho\partial_\rho$ is the d'Alembert operator, and $\kappa=8\pi G/c^4$ is Einstein's constant, where $G$ is the gravitational constant. The coefficients on the left in Eq.~\eqref{eq:UGgravity} are given by $P^{\mu\nu,\rho\sigma}=\frac{1}{2}(\eta^{\mu\sigma}\eta^{\rho\nu}+\eta^{\mu\rho}\eta^{\nu\sigma}-\eta^{\mu\nu}\eta^{\rho\sigma})$. The source of the gravitational field is the stress-energy-momentum tensor, $T_\mathrm{m}^{\mu\nu}$, which appears on the right in Eq.~\eqref{eq:UGgravity}. The stress-energy-momentum tensor describes the energy, momentum, and angular momentum content of the fields other than the gravitational field. The stress-energy-momentum tensor of a classical point mass $M$ located at the origin is well-known to be given by \cite{Landau1989,Partanen2025c}
\begin{equation}
 T_\mathrm{m}^{\mu\nu}=Mc^2\delta(\mathbf{r})\delta_0^\mu\delta_0^\nu.
 \label{eq:SEM}
\end{equation}
Here $\delta(\mathbf{r})$ is the three-dimensional Dirac delta function, $\mathbf{r}=(x,y,z)$ denotes the three-dimensional space coordinates, and $\delta_\nu^\mu$ is the Kronecker delta.

Using the stress-energy-momentum tensor source term in Eq.~\eqref{eq:SEM}, the solution of Eq.~\eqref{eq:UGgravity} for the gravity gauge field $H_{\mu\nu}$ is given by \cite{Partanen2025c}
\begin{equation}
 H_{\mu\nu}=\left[
 \begin{array}{cccc}
 \frac{\Phi}{c^2} & 0 & 0 & 0\\
 0 & \frac{\Phi}{c^2} & 0 & 0\\
 0 & 0 & \frac{\Phi}{c^2} & 0\\
 0 & 0 & 0 & \frac{\Phi}{c^2}
 \end{array}\right],\hspace{0.5cm}\Phi=-\frac{GM}{r}.
 \label{eq:H}
\end{equation}
Here $\Phi$ is the Newtonian gravitational potential, where $r=|\mathbf{r}|=\sqrt{x^2+y^2+z^2}$. The integration constants have been set to zero by assuming that the gravitational field vanishes at infinity. The gravitational potential satisfies Poisson's equation $\nabla^2\Phi=4\pi GM\delta(\mathbf{r})$, where $\nabla=(\partial_x,\partial_y,\partial_z)$ is the three-dimensional vector differential operator.

\section{Electric potential of the atomic nucleus}
\vspace{-0.3cm}

The dynamical equation of the electromagnetic four-potential $A^\mu$ in UG is given in the Feynman gauge, $\partial_\mu A^\mu=0$, by \cite{Partanen2025a,Partanen2025b}
\begin{align}
 \partial^2A^\sigma+P^{\mu\nu,\rho\sigma,\eta\lambda}\partial_\rho(H_{\mu\nu}\partial_\eta A_\lambda) &=\mu_0J_\mathrm{e,tot}^\sigma.
 \label{eq:MaxwellUG}
\end{align}
Here $\mu_0$ is the permeability of vacuum, and the coefficients $P^{\mu\nu,\rho\sigma,\eta\lambda}$ are given in terms of the Minkowski metric by \cite{Partanen2025a,Partanen2025c}
\begin{align}
 &P^{\mu\nu,\rho\sigma,\eta\lambda}
 =\eta^{\eta\sigma}\eta^{\lambda\mu}\eta^{\nu\rho}-\eta^{\eta\mu}\eta^{\lambda\sigma}\eta^{\nu\rho}-\eta^{\eta\rho}\eta^{\lambda\mu}\eta^{\nu\sigma}\nonumber\\
 &\hspace{0.3cm}+\eta^{\eta\mu}\eta^{\lambda\rho}\eta^{\nu\sigma}-\eta^{\mu\sigma}\eta^{\nu\lambda}\eta^{\rho\eta}+\eta^{\mu\sigma}\eta^{\nu\eta}\eta^{\rho\lambda}+\eta^{\mu\rho}\eta^{\nu\lambda}\eta^{\sigma\eta}\nonumber\\
 &\hspace{0.3cm}-\eta^{\mu\rho}\eta^{\nu\eta}\eta^{\sigma\lambda}-\eta^{\mu\nu}\eta^{\eta\sigma}\eta^{\lambda\rho}+\eta^{\mu\nu}\eta^{\eta\rho}\eta^{\lambda\sigma}.
 \label{eq:P3}
\end{align}

The total electric four-current density of UG is denoted by $J_\mathrm{e,tot}^\mu=J_\mathrm{e}^\rho-P^{\mu\nu,\rho\sigma}J_\mathrm{e\sigma}H_{\mu\nu}$, where $J_\mathrm{e}^\rho$ is given by $J_\mathrm{e}^\rho=q_{\mathrm{e}i}c\bar{\psi}_i\boldsymbol{\gamma}^\rho\psi_i$, where $\psi_i$ are the Dirac fields of the theory, indexed by $i$, and $q_{\mathrm{e}i}$ are the pertinent charges. The conserved current property of $J_\mathrm{e,tot}^\mu$ is discussed in Methods. In the following, we use Eq.~\eqref{eq:MaxwellUG} to derive the four-potential of the atomic nucleus to be later used in the Dirac equation of the electrons. Therefore, in the total electric four-current density, we include here only the contribution of the atomic nucleus, which is approximated by a delta function. Thus, the electric four-current density of the nucleus of a hydrogen-like atom, containing $Z$ protons with total charge $Ze$ at rest at position $\mathbf{r}=\mathbf{r}_0$, is given by
\begin{equation}
 J_\mathrm{e,tot}^\mu=Zec\delta_0^\mu\delta(\mathbf{r}-\mathbf{r}_0).
 \label{eq:J}
\end{equation}

Assuming the electromagnetic four-potential of the form $A^\mu=(\phi_\mathrm{e}/c,0,0,0)$, where $\phi_\mathrm{e}$ is the electric scalar potential that is independent of time, the terms on both sides of Eq.~\eqref{eq:MaxwellUG} with $\sigma\in\{x,y,z\}$ are identically zero. For the component of Eq.~\eqref{eq:MaxwellUG} with $\sigma=0$, we obtain
\begin{equation}
 \Big(1-\frac{2\Phi}{c^2}\Big)\nabla^2\phi_\mathrm{e}
 -\frac{2}{c^2}\nabla\Phi\cdot\nabla\phi_\mathrm{e}
 =-\frac{Ze}{\varepsilon_0}\delta(\mathbf{r}-\mathbf{r}_0).
 \label{eq:MaxwellUG2}
\end{equation}
Here $\varepsilon_0=1/(\mu_0c^2)$ is the permittivity of vacuum. The derivatives of the very smoothly varying Newtonian potential $\Phi$ are exceedingly small in comparison with the derivatives of the electromagnetic potential $\phi_\mathrm{e}$ of the atomic nucleus. Therefore, we can approximate $\Phi$ by the constant term of its Taylor series at $\mathbf{r}=\mathbf{r}_0$, denoted by $\Phi_0$. Thus, writing $\Phi=\Phi_0$, dropping out the second term on the left in Eq.~\eqref{eq:MaxwellUG2} and dividing the resulting equation by $1-2\Phi_0/c^2$, we obtain
\begin{equation}
 \nabla^2\phi_\mathrm{e}=-\frac{Ze}{\varepsilon_0\big(1-\frac{2\Phi_0}{c^2}\big)}\delta(\mathbf{r}-\mathbf{r}_0).
 \label{eq:MaxwellUG3}
\end{equation}
Apart from the prefactor, Eq.~\eqref{eq:MaxwellUG3} is the well-known equation defining the Coulomb potential \cite{Jackson1999}. Therefore, the solution to Eq.~\eqref{eq:MaxwellUG3} is given by
\begin{equation}
 \phi_\mathrm{e}=\frac{Ze}{4\pi\varepsilon_0\big(1-\frac{2\Phi_0}{c^2}\big)|\mathbf{r}-\mathbf{r}_0|}.
 \label{eq:Ve}
\end{equation}

\section{Dirac equation of electrons}
\vspace{-0.3cm}

Next, we study the electron eigenstates of hydrogen-like atoms using the Dirac equation in the presence of the gravity gauge field, $H_{\mu\nu}$ of Eq.~\eqref{eq:H}, and the electromagnetic four-potential of the atomic nucleus, $A_\mu$ defined through Eqs.~\eqref{eq:MaxwellUG}--\eqref{eq:Ve}. The Dirac equation of UG is written for an electron with mass $m_\mathrm{e}$ and charge $-e$ as \cite{Partanen2025a}
\begin{align}
 &i\hbar c\boldsymbol{\gamma}^\rho\vec{\partial}_\rho\psi\!-\!m_\mathrm{e}c^2\psi
 \!=-ec\boldsymbol{\gamma}^\rho\psi A_\rho
 +P^{\mu\nu,\rho\sigma}\Big(i\hbar c\boldsymbol{\gamma}_\sigma\vec{\partial}_\rho\psi\nonumber\\
 &\hspace{0.4cm}-\frac{m_\mathrm{e}c^2}{2}\eta_{\rho\sigma}\psi+\frac{i\hbar c}{2}\boldsymbol{\gamma}_\sigma\psi\vec{\partial}_\rho
 +ec\boldsymbol{\gamma}_\sigma\psi A_\rho
 \Big)H_{\mu\nu}.
 \label{eq:DiracUGM}
\end{align}
Here $\hbar$ is the reduced Planck constant, $\psi$ is the Dirac field describing the electron, and $\boldsymbol{\gamma}^\mu$ are the conventional $4\times 4$ Dirac gamma matrices.

Substituting the gravity gauge field $H_{\mu\nu}$ from  Eq.~\eqref{eq:H} into Eq.~\eqref{eq:DiracUGM}, using the electromagnetic four-potential of the nucleus, $A^\mu=(\phi_\mathrm{e}/c,0,0,0)$, and approximating $\Phi=\Phi_0$, we obtain
\begin{align}
 &\Big(1-\frac{\Phi_0}{c^2}\Big)m_\mathrm{e}c^2\boldsymbol{\beta}\psi
 +c\boldsymbol{\alpha}\cdot\hat{\mathbf{p}}\psi
 -\Big(1-\frac{2\Phi_0}{c^2}\Big)e\phi_\mathrm{e}\psi\nonumber\\
 &=i\hbar\Big(1-\frac{2\Phi_0}{c^2}\Big)\frac{\partial \psi}{\partial t}.
 \label{eq:DiracUGM4}
\end{align}
Here $\hat{\mathbf{p}}=-i\hbar\nabla$ is the three-dimensional momentum operator. The alpha and beta matrices of the Dirac theory are given in terms of the gamma matrices by $\boldsymbol{\beta}=\boldsymbol{\gamma}^0$ and $\boldsymbol{\alpha}^i=\boldsymbol{\gamma}^0\boldsymbol{\gamma}^i$, $i\in\{x,y,z\}$, and the three-component alpha matrix vector is given by $\boldsymbol{\alpha}=(\boldsymbol{\alpha}^x,\boldsymbol{\alpha}^y,\boldsymbol{\alpha}^z)$.

Using the Dirac equation in Eq.~\eqref{eq:DiracUGM4} and the electric potential of the atomic nucleus in Eq.~\eqref{eq:Ve}, and dividing Eq.~\eqref{eq:DiracUGM4} by $1-2\Phi_0/c^2$, we can then rewrite the Dirac equation of UG in the Hamiltonian form, given by
\begin{equation}
 \begin{array}{c}
 \hat{H}\psi=i\hbar\dfrac{\partial \psi}{\partial t},
 \hspace{0.5cm}\hat{H}=C_1m_\mathrm{e}c^2\boldsymbol{\beta}+C_2c\boldsymbol{\alpha}\cdot\hat{\mathbf{p}}-C_2\dfrac{Z\hbar c\alpha_\mathrm{e}}{|\mathbf{r}-\mathbf{r}_0|},\\[10pt]
 C_1=\dfrac{1-\frac{\Phi_0}{c^2}}{1-\frac{2\Phi_0}{c^2}},\hspace{0.5cm}C_2=\dfrac{1}{1-\frac{2\Phi_0}{c^2}}.
 \end{array}
 \label{eq:Hamiltonian}
\end{equation}
Here $\alpha_\mathrm{e}=e^2/(4\pi\varepsilon_0\hbar c)$ is the electromagnetic fine-structure constant, $\hat{H}$ is the Hamiltonian operator, and we have defined the constant coefficients $C_1$ and $C_2$ to contain the dependencies on the Newtonian potential. The conventional Dirac Hamiltonian in the electric potential of the atomic nucleus is recovered at zero gravitational potential, for which the values of $C_1$ and $C_2$ are equal to unity. In the general case, the non-unity values of these coefficients have fundamental consequences, such as the gravitational redshift obtained below.

\section{Electron eigenstates}
\vspace{-0.3cm}

We observe that Eq.~\eqref{eq:Hamiltonian} can be rewritten in the conventional form of the Dirac equation in the electric potential by using quantities scaled by $C_1$ and $C_2$. From the known solution of the Dirac equation for the hydrogen-like atoms in QED \cite{Grant2007,Landau1982}, we can obtain the solution in the gravitational potential by simply replacing the quantities $m_\mathrm{e}$ and $c$ in the conventional solution by $(C_1/C_2^2)m_\mathrm{e}$ and $C_2c$, respectively. Therefore, denoting $\mathbf{r}'=\mathbf{r}-\mathbf{r}_0$, and using spherical coordinates, in which $\mathbf{r}'=(r',\theta',\phi')$, the eigenstates of the Dirac equation of UG in Eq.~\eqref{eq:Hamiltonian} are given by
\begin{align}
 &\psi_{n_\mathrm{r},\kappa_\mathrm{r},j,m}(t,r',\theta',\phi')\nonumber\\
 &=\theta(\kappa_\mathrm{r})\left[\begin{array}{c}
  f_{n_\mathrm{r},\kappa_\mathrm{r}}(r')\Omega_{j,j+\frac{1}{2},m}(\theta',\phi')\\
  ig_{n_\mathrm{r},\kappa_\mathrm{r}}(r')\Omega_{j,j-\frac{1}{2},m}(\theta',\phi')
  \end{array}\right]e^{-iE_{n_\mathrm{r},\kappa_\mathrm{r}}t/\hbar}\nonumber\\
  &\hspace{0.4cm}+\theta(-\kappa_\mathrm{r})\left[\begin{array}{c}
  f_{n_\mathrm{r},\kappa_\mathrm{r}}(r')\Omega_{j,j-\frac{1}{2},m}(\theta',\phi')\\
  ig_{n_\mathrm{r},\kappa_\mathrm{r}}(r')\Omega_{j,j+\frac{1}{2},m}(\theta',\phi')
  \end{array}\right]e^{-iE_{n_\mathrm{r},\kappa_\mathrm{r}}t/\hbar}.
  \label{eq:psi}
\end{align}
Here $\theta(x)$ is the unit step function, $\Omega_{j,l,m}(\theta,\phi)$ are the spherical harmonic spinors, and $f_{n_\mathrm{r},\kappa_\mathrm{r}}(r)$ and $g_{n_\mathrm{r},\kappa_\mathrm{r}}(r)$ are the radial functions. The radial functions depend on the coefficients $C_1$ and $C_2$, while the energies $E_{n_\mathrm{r},\kappa_\mathrm{r}}$ depend on the coefficient $C_1$ only. For the explicit expressions of the spherical harmonic spinors and the radial functions, see Methods.
The energies of the electron eigenstates are given by
\begin{equation}
 \begin{array}{c}
 E_{n_\mathrm{r},\kappa_\mathrm{r}}=C_1E_{n_\mathrm{r},\kappa_\mathrm{r}}^{(0)},\\[10pt]
 E_{n_\mathrm{r},\kappa_\mathrm{r}}^{(0)}=m_\mathrm{e}c^2\Bigg(1+\dfrac{(Z\alpha_\mathrm{e})^2}{\big(n_\mathrm{r}+\sqrt{\kappa_\mathrm{r}^2-(Z\alpha_\mathrm{e})^2}\big)^2}\Bigg)^{-1/2}.
 \end{array}
 \label{eq:energy}
\end{equation}
Here $E_{n_\mathrm{r},\kappa_\mathrm{r}}^{(0)}$ are the well-known energies of the Dirac eigenstates for hydrogen-like atoms in the absence of the gravitational field. The quantum numbers of the Dirac eigenstates are conventional and described in Methods.

\section{Gravitational redshift}
\vspace{-0.3cm}

Equation \eqref{eq:energy} shows that the energies of all electron eigenstates are scaled by the factor $C_1$. Therefore, the frequencies of the emitted photons are scaled by the same factor. The frequency of photons does not change under propagation in the global Minkowski frame. Therefore, the frequency of the emitted photon calculated at the position of the atom in the global Minkowski frame is the same as the frequency of the photon detected after propagation at zero gravitational potential. Here we assume that the atom and the detector are not moving with respect to each other, whence the relativistic Doppler shifts are absent. When the angular frequency of the photon emitted by the atom at zero gravitational potential is denoted by $\omega_\mathrm{e}$ and the frequency of the photon emitted at nonzero gravitational potential is denoted by $\omega_\mathrm{r}$, these frequencies are related by
\begin{equation}
 \begin{array}{c}
 \omega_\mathrm{r}=C_1\omega_\mathrm{e},\\[10pt]
 \omega_\mathrm{e}=\frac{1}{\hbar}(E_{n_\mathrm{r1},\kappa_\mathrm{r1}}^{(0)}-E_{n_\mathrm{r2},\kappa_\mathrm{r2}}^{(0)}).
 \end{array}
 \label{eq:frequency}
\end{equation}
Here $n_\mathrm{r1}$ and $\kappa_\mathrm{r1}$ are the quantum numbers of the initial atomic state and $n_\mathrm{r2}$ and $\kappa_\mathrm{r2}$ are the quantum numbers of the final atomic state. Accordingly, the gravitational redshift is given by
\begin{equation}
 z_\mathrm{UG}=\frac{\omega_\mathrm{e}-\omega_\mathrm{r}}{\omega_\mathrm{r}}=\frac{1}{C_1}-1
 \approx\frac{GM}{r_0c^2}-\Big(\frac{GM}{r_0c^2}\Big)^2.
 \label{eq:redshiftUG}
\end{equation}
Here $r_0=|\mathbf{r}_0|$ is the distance between the emitting atom and the center of the gravitational potential in the global Minkowski frame. The last form of Eq.~\eqref{eq:redshiftUG} is obtained by expanding the Taylor series in powers of $GM/(r_0c^2)$ and truncating it after the second-order term.

The corresponding result of GR up to the second-order term in the gravitational constant is given by \cite{Misner1973,Moore2013}
\begin{equation}
 z_\mathrm{GR}
 =\frac{\sqrt{g_{00}(\text{receiver})}}{\sqrt{g_{00}(\text{emitter})}}-1
 \approx\frac{GM}{r_0c^2}+\frac{1}{2}\Big(\frac{GM}{r_0c^2}\Big)^2.
 \label{eq:redshiftGR}
\end{equation}
The last form of Eq.~\eqref{eq:redshiftGR} is obtained by using the time-time component of the Schwarzschild metric in the isotropic coordinates at the emitter and receiver positions. We have then taken the first few terms of the pertinent Taylor series. The first-order term obtained from UG in Eq.~\eqref{eq:redshiftUG} is equivalent to that obtained from GR in Eq.~\eqref{eq:redshiftGR} \cite{Misner1973,Moore2013}. However, the theories differ regarding the higher-order corrections, which have not been experimentally measured yet \cite{Zheng2023,Pumpo2021,Ufrecht2020,GonzalezHernandez2020,Herrmann2018,Delva2018,Wolf2016,Kraniotis2021}.

\section{Symmetry breaking by the gravitational potential gradient}

Next, we briefly discuss accounting for the gravitational potential gradient, which was neglected when writing the equation for the electric potential of the atomic nucleus in Eq.~\eqref{eq:MaxwellUG3}. Using the first two terms in the Taylor series of $\Phi$ by writing $\Phi=\Phi_0+\nabla\Phi|_{\mathbf{r}=\mathbf{r}_0}\cdot(\mathbf{r}-\mathbf{r}_0)$ and dividing Eq.~\eqref{eq:MaxwellUG3} by $1-2\Phi/c^2\approx 1-2\Phi_0/c^2$, we obtain
\begin{align}
\begin{array}{c}
 \nabla^2\phi_\mathrm{e}
 -\mathbf{a}\cdot\nabla\phi_\mathrm{e}
 =-\dfrac{Ze}{\varepsilon_0\big(1-\frac{2\Phi_0}{c^2}\big)}\delta(\mathbf{r}-\mathbf{r}_0),\\[10pt]
 \mathbf{a}=\dfrac{2\nabla\Phi|_{\mathbf{r}=\mathbf{r}_0}}{c^2\big(1-\frac{2\Phi_0}{c^2}\big)}=\dfrac{2GM\mathbf{r}_0}{c^2r_0^3\big(1+\frac{2GM}{r_0c^2}\big)}.
\end{array}
 \label{eq:DerivativeEffects1}
\end{align}
The constant vector $\mathbf{a}$ in this approximation modifies Eq.~\eqref{eq:MaxwellUG3}, and, consequently, it changes the electric potential solution in Eq.~\eqref{eq:Ve} to become
\begin{align}
 &\phi_\mathrm{e} =\frac{Ze\exp[\frac{1}{2}\mathbf{a}\cdot(\mathbf{r}-\mathbf{r}_0)+\frac{1}{2}|\mathbf{a}||\mathbf{r}-\mathbf{r}_0|]}{4\pi\varepsilon_0\big(1-\frac{2\Phi_0}{c^2}\big)|\mathbf{r}-\mathbf{r}_0|}\nonumber\\
 &\approx\frac{Ze}{4\pi\varepsilon_0\big(1-\frac{2\Phi_0}{c^2}\big)|\mathbf{r}-\mathbf{r}_0|}
 +\frac{Ze[\mathbf{a}\cdot(\mathbf{r}-\mathbf{r}_0)+|\mathbf{a}||\mathbf{r}-\mathbf{r}_0|]}{8\pi\varepsilon_0\big(1-\frac{2\Phi_0}{c^2}\big)|\mathbf{r}-\mathbf{r}_0|}.
 \label{eq:VeMod}
\end{align}
The last form of Eq.~\eqref{eq:VeMod} is obtained by taking the first two terms of the Taylor series of the exponential function at $\mathbf{r}=\mathbf{r}_0$. The last term of Eq.~\eqref{eq:VeMod} represents a small perturbation to the electric potential in Eq.~\eqref{eq:Ve}. Thus, it could be accounted for in the solution of the Dirac equation by using perturbation theory. We conclude that the last term of Eq.~\eqref{eq:Ve} breaks the spherical symmetry of the electric potential of the atomic nucleus. Therefore, it is expected to lead to splitting of spectral lines of certain atomic states that are otherwise degenerate in analogy with the effect of the external magnetic field in the Zeeman effect of QED \cite{Landau1982,Grant2007}. This enables further interesting tests for UG in the presence of strong gravitational potential gradients. The present work enables detailed spectral line analysis, which, however, is left as a topic of further work.

\section{Conclusion}
\vspace{-0.3cm}

We have shown how the quantum-field-theory-based calculation within UG explains gravitational redshift in a fundamentally different way in comparison with the metric-based calculation of GR. Using the Dirac equation of UG, we have calculated the shift of atomic energy levels in the gravity gauge field and found agreement with the experimentally observable gravitational redshift. The gravitational redshifts obtained from UG and GR agree within the first-order term in the gravitational constant, but higher-order terms lead to differences. The higher-order terms have not been experimentally measured yet, and the difference of the theories can be tested by more precise experiments in the future. In addition to gravitational redshift \cite{Zheng2023,Pumpo2021,Ufrecht2020,GonzalezHernandez2020,Herrmann2018,Delva2018,Wolf2016,Kraniotis2021}, the theories can also be compared based on their predictions for gravitational lensing \cite{Will2014,Turyshev2007,Turyshev2009b,Ulbricht2020} and gravitational wave data \cite{Abbott2016a,Abbott2016b,Abbott2017}. The new experiments could decide if the underlying assumptions limit GR to the first order in the gravitational constant, and it is equally interesting if UG agrees with the experiments to the second order. Furthermore, strong gravitational potential gradients in UG are expected to split certain spectral lines by breaking the spherical symmetry of the electric potential of the atomic nucleus. Since UG includes the full quantum description of the field-matter interaction in the gravitational field, it can be used to analyze all classical and quantum photonics phenomena in astrophysical optics.




\appendix

\section*{Methods}

\section{\label{apx:Lagrangian}Lagrangian density}
\vspace{-0.3cm}

Using the geometric condition corresponding to the Minkowski spacetime together with the equivalence principles of scale and mass, the gauge-fixed Lagrangian density of UG in the absence of strong and weak interactions is given by \cite{Partanen2025a}
\begin{align}
 \mathcal{L}&=\frac{i\hbar c}{2}\bar{\psi}_i(\boldsymbol{\gamma}^\nu\vec{\partial}_\nu-\cev{\partial}_\nu\boldsymbol{\gamma}^\nu)\psi_i-m_ic^2\bar{\psi}_i\psi_i
 -\frac{1}{4\mu_0}F_{\mu\nu}F^{\mu\nu}\nonumber\\
 &\hspace{0.4cm}+\frac{1}{8\kappa}(H_{\rho\mu\nu}H^{\rho\mu\nu}+2H_{\rho\mu\nu}H^{\mu\rho\nu}-4H_{\;\,\mu\nu}^\nu H_{\;\;\;\,\rho}^{\rho\mu})\nonumber\\
 &\hspace{0.4cm}-J_\mathrm{e}^\nu A_\nu-T_\mathrm{m}^{\mu\nu}H_{\mu\nu}
 -\frac{1}{2\mu_0\xi_\mathrm{e}}(\partial_{\nu}A^{\nu})^2\nonumber\\
 &\hspace{0.4cm}+\frac{1}{\kappa\xi_\mathrm{g}}\eta_{\gamma\delta}P^{\alpha\beta,\lambda\gamma}P^{\rho\sigma,\eta\delta}\partial_\lambda H_{\alpha\beta}\partial_\eta H_{\rho\sigma}.
 \label{eq:Lagrangian}
\end{align}
The implicit summation over $i$ is over all fermion fields of the theory. In Eq.~\eqref{eq:Lagrangian}, $m_i$ is the mass of the fermion of index $i$, $\xi_\mathrm{e}$ and $\xi_\mathrm{g}$ are the electromagnetic and gravity gauge fixing parameters, $F_{\mu\nu}$ and $H_{\rho\mu\nu}$ are the electromagnetic and gravity field-strength tensors, $J_\mathrm{e}^\nu$ is the electric four-current density, and $T_\mathrm{m}^{\mu\nu}$ is the stress-energy-momentum tensor of the electromagnetic and fermion fields. These quantities are given by \cite{Partanen2025a,Partanen2025b}
\begin{equation}
 F_{\mu\nu}=\partial_\mu A_\nu-\partial_\nu A_\mu,
\end{equation}
\begin{equation}
 H_{\rho\mu\nu}=\partial_\mu H_{\rho\nu}-\partial_\nu H_{\rho\mu},
\end{equation}
\begin{equation}
 J_\mathrm{e}^\nu=q_{\mathrm{e}i}c\bar{\psi}_i\boldsymbol{\gamma}^\nu\psi_i,
 \label{eq:Je}
\end{equation}
\begin{align}
 T_\mathrm{m}^{\mu\nu} &=\frac{c}{2}P^{\mu\nu,\rho\sigma}[i\hbar\bar{\psi}_i(\boldsymbol{\gamma}_\mathrm{\rho}\vec{\partial}_\sigma-\cev{\partial}_\rho\boldsymbol{\gamma}_\mathrm{\sigma})\psi_i\nonumber\\
 &\hspace{0.4cm}-q_\mathrm{e}\bar{\psi}_i(\boldsymbol{\gamma}_\mathrm{\rho}A_\sigma+A_\rho\boldsymbol{\gamma}_\mathrm{\sigma})\psi_i
 -m_\mathrm{e}c\eta_{\rho\sigma}\bar{\psi}_i\psi_i]\nonumber\\
 &\hspace{0.4cm}+\frac{1}{2\mu_0}P^{\mu\nu,\rho\sigma,\eta\lambda}\partial_\rho A_\sigma\partial_\eta A_\lambda.
\end{align}
Here $q_{\mathrm{e}i}$ is the electric charge of the fermion field of index $i$. Note that, in the presence of the gravitational field, $J_\mathrm{e}^\nu$ is not the total electric four-current density of UG as discussed below.

\section{\label{apx:J}Electric four-current density}
\vspace{-0.3cm}

The electric four-current density $J_\mathrm{e}^\nu$ in Eq.~\eqref{eq:Je} is not a conserved quantity in UG in the presence of the gravitational field. Instead, the conserved current associated with the electromagnetic gauge field is calculated as presented below.

The unitary transformation associated with the U(1) gauge symmetry of QED is given by
\begin{equation}
 \psi_i\rightarrow U_\mathrm{e}\psi_i,\hspace{0.5cm}\text{where }U_\mathrm{e}=e^{i\theta Q}.
 \label{eq:U1transformation}
\end{equation}
Here $\theta$ is the real-valued symmetry transformation parameter, and $Q$ is the symmetry transformation generator, which has the value of $Q=-1$ for electrons. The infinitesimal variation of the Dirac field $\psi_i$, in the symmetry transformation of Eq.~\eqref{eq:U1transformation} with respect to the symmetry transformation parameter $\theta$ is given by
\begin{equation}
 \delta\psi_i=iQ_i\psi_i\delta\theta.
 \label{eq:deltapsi}
\end{equation}
Here $Q_i$ is the charge quantum number for the fermion field of index $i$.

Using the infinitesimal variation of the Dirac field in Eq.~\eqref{eq:deltapsi} at zero electromagnetic gauge field, $A^\mu=0$, the variation of the Lagrangian density of UG in Eq.~\eqref{eq:Lagrangian} is written as
\begin{align}
 &\delta\mathcal{L}|_{A=0}\nonumber\\
 &=\frac{i\hbar c}{2}(\delta\bar{\psi}_i)(\boldsymbol{\gamma}^\rho\vec{\partial}_\rho-\cev{\partial}_\rho\boldsymbol{\gamma}^\rho)\psi_i
 +\frac{i\hbar c}{2}\bar{\psi}_i(\boldsymbol{\gamma}^\rho\vec{\partial}_\rho-\cev{\partial}_\rho\boldsymbol{\gamma}^\rho)(\delta\psi_i)\nonumber\\
 &\hspace{0.5cm}-m_\mathrm{e}c^2(\delta\bar{\psi}_i)\psi_i
 -m_\mathrm{e}c^2\bar{\psi}_i(\delta\psi_i)
 -\frac{c}{2}P^{\mu\nu,\rho\sigma}\nonumber\\
 &\hspace{0.5cm}\times[i\hbar(\delta\bar{\psi}_i)(\boldsymbol{\gamma}_\mathrm{\rho}\vec{\partial}_\sigma\!-\!\cev{\partial}_\rho\boldsymbol{\gamma}_\sigma)\psi_i
 \!+\!i\hbar\bar{\psi}_i(\boldsymbol{\gamma}_\mathrm{\rho}\vec{\partial}_\sigma\!-\!\cev{\partial}_\rho\boldsymbol{\gamma}_\sigma)(\delta\psi_i)\nonumber\\
 &\hspace{0.5cm}-m_ic\eta_{\rho\sigma}(\delta\bar{\psi}_i)\psi_i-m_ic\eta_{\rho\sigma}\bar{\psi}_i(\delta\psi_i)]H_{\mu\nu}\nonumber\\
 &=-\hbar Q_ic\bar{\psi}_i(\boldsymbol{\gamma}^\rho
 -P^{\mu\nu,\rho\sigma}\boldsymbol{\gamma}_\sigma H_{\mu\nu})\psi_i\partial_\rho\delta\theta\nonumber\\
 &=-\frac{\hbar}{e}J_\mathrm{e,tot}^\rho\partial_\rho\delta\theta.
 \label{eq:Lvariation}
\end{align}
The last equality of Eq.~\eqref{eq:Lvariation} defines the total conserved electric four-current density of UG, given by
\begin{align}
 J_\mathrm{e,tot}^\rho
 &=q_{\mathrm{e}i}c\bar{\psi}_i(\boldsymbol{\gamma}^\rho
 -P^{\mu\nu,\rho\sigma}\boldsymbol{\gamma}_\sigma H_{\mu\nu})\psi_i\nonumber\\
 &=J_\mathrm{e}^\rho-P^{\mu\nu,\rho\sigma}J_\mathrm{e\sigma}H_{\mu\nu}.
 \label{eq:Jtot}
\end{align}
Here $q_{\mathrm{e}i}=Q_ie$ is the electric charge of the particle. In the last equality of Eq.~\eqref{eq:Jtot}, we have used the conventional definition of the electric four-current density in the absence of the gravity gauge field, given in Eq.~\eqref{eq:Je}.

Substituting the gravity gauge field from Eq.~\eqref{eq:H} into Eq.~\eqref{eq:Jtot}, we obtain after technical summation over repeated indices
\begin{equation}
 J_\mathrm{e,tot}^\rho=J_\mathrm{e}^\rho-\frac{2\Phi}{c^2}\delta_0^\rho J_\mathrm{e}^0.
 \label{eq:Jtot2}
\end{equation}
In the classical limit in the rest frame of the particle, the total conserved electric four-current density in Eq.~\eqref{eq:Jtot2} can be approximated by the Dirac delta function as
\begin{equation}
 J_\mathrm{e,tot}^\rho=q_\mathrm{e}c\delta_0^\rho\delta(\mathbf{r}-\mathbf{r}_0).
 \label{eq:Jtot3}
\end{equation}
The conserved charge is then obtained by integration of $J_\mathrm{e,tot}^0$ over the volume as $\int J_\mathrm{e,tot}^0d^3r=q_\mathrm{e}c$. Using Eq.~\eqref{eq:Jtot3}, we can solve Eq.~\eqref{eq:Jtot2} for $J_\mathrm{e}^\rho$ as
\begin{equation}
 J_\mathrm{e}^\rho=\frac{q_\mathrm{e}c}{1-\frac{2\Phi_0}{c^2}}\delta_0^\rho\delta(\mathbf{r}-\mathbf{r}_0).
 \label{eq:Je2}
\end{equation}
The quantity $J_\mathrm{e}^\rho$ is not the conserved current since, for a particle at rest, the integral of $J_\mathrm{e}^0$ over the volume depends on the position through the Newtonian potential $\Phi_0$ in the prefactor of Eq.~\eqref{eq:Je2}.

Next, we derive the conservation law of the total electric four-current density of UG. The variation of the action integral, i.e., the integral of the Lagrangian density in Eq.~\eqref{eq:Lagrangian} over the volume, with respect to $\theta$ is given for zero electromagnetic four-potential, $A^\mu=0$, by
\begin{align}
 \delta S|_{A=0} &=\int\delta\mathcal{L}|_{A=0}d^4x
 =-\int\frac{\,\hbar}{e}J_\mathrm{e,tot}^\nu\partial_\nu\delta\theta d^4x\nonumber\\
 &=-\int\partial_\nu\Big(\frac{\,\hbar}{e}J_\mathrm{e,tot}^\nu\delta\theta\Big)d^4x
 +\int\partial_\nu\Big(\frac{\hbar}{e}J_\mathrm{e,tot}^\nu\Big)\delta\theta d^4x\nonumber\\
 &=\int\frac{\hbar}{e}\partial_\nu J_\mathrm{e,tot}^\nu\delta\theta d^4x.
 \label{eq:Jeconservation}
\end{align}
In the second equality of Eq.~\eqref{eq:Jeconservation}, we have used Eq.~\eqref{eq:Lvariation}. In the third equality, we have applied integration by parts. In the fourth equality, we have dropped out the total divergence term, which is zero when the fields vanish at the distant boundary. The result of Eq.~\eqref{eq:Jeconservation} shows that the variation of the action integral vanishes for arbitrary $\delta\theta$ when
\begin{equation}
 \partial_\nu J_\mathrm{e,tot}^{\nu}=0.
 \label{eq:Jeconservation1}
\end{equation}
This is the well-known form of the conservation law of the total electric four-current density in the Cartesian Minkowski spacetime \cite{Jackson1999}.

\section{\label{apx:representation}Quantum numbers, spherical harmonic spinors, and radial functions}
\vspace{-0.3cm}

In the eigenstates of the Dirac equation in Eq.~\eqref{eq:psi}, the quantity $j=l\pm\frac{1}{2}$ is the total angular momentum quantum number, where $l=0,1,\ldots,n-1$ is the orbital angular momentum quantum number. The magnetic quantum number $m=-j,-j+1,\ldots,j$ is the total angular momentum projection onto the $z$-axis. The relativistic angular quantum number $\kappa_\mathrm{r}=\pm(j+\frac{1}{2})$ takes all integer values except zero. The positive values of $\kappa_\mathrm{r}$ correspond to the case $j=l-\frac{1}{2}$, and the negative values to the case $j=l+\frac{1}{2}$. The principal quantum number is given by $n=n_\mathrm{r}+j+\frac{1}{2}$. The radial quantum number $n_\mathrm{r}$ takes integer values $n_\mathrm{r}=0,1,2,\ldots$ for $\kappa_\mathrm{r}<0$ and $n_\mathrm{r}=1,2,3,\ldots$ for $\kappa_\mathrm{r}>0$.

In the representation of the Dirac equation eigenstates in Eq.~\eqref{eq:psi}, the functions $\Omega_{j,l,m}(\theta,\phi)$ are the spherical harmonic spinors, defined as \cite{Grant2007,Landau1982}
\begin{equation}
 \Omega_{j,l,m}(\theta,\phi)
 =\sum_{q=-\frac{1}{2}}^{\frac{1}{2}}\langle l,m-q,\textstyle\frac{1}{2},q|j,m\rangle 
Y_{l,m-q}(\theta,\phi)u^{(q)}.
\end{equation}
The terms of this series are formed from the well-known Clebsch-Gordan coefficients $\langle j_1,m_1,j_2,m_2|j_3,m_3\rangle$, the scalar spherical harmonic functions $Y_{l,m}(\theta,\phi)$, and the spherical unit spinors $u^{(q)}$. The spherical unit spinors $u^{(q)}$ are defined as $u^{(-1/2)}=(0,1)$ and $u^{(1/2)}=(1,0)$. For the scalar spherical harmonic functions, we use the definition written in terms of the associated Legendre polynomials $P_{l,m}(x)$ as
\begin{equation} 
Y_{l,m}(\theta,\phi)=\sqrt{\frac{2l+1}{4\pi}\frac{
(l-m)!}{(l+m)!}}\,P_{l,m}(\cos\theta)e^{im\phi}.
\end{equation}
For the associated Legendre polynomials, we use the Condon-Shortley phase 
convention. The associated Legendre polynomials $P_{l,m}(x)$ are then given by
\begin{equation}
 P_{l,m}(x)=
\left\{\begin{array}{l}
(-1)^m(1-x^2)^{m/2}\dfrac{d^m}{dx^m}P_l(x),\hspace{0.5cm}m\ge0,\\
(-1)^{-m}\dfrac{(l+m)!}{(l-m)!}P_{l,-m}(x),\hspace{0.5cm}m<0.      
\end{array}\right.
\end{equation}
Here $P_l(x)$ is the conventional Legendre polynomial of degree $l$, given by the Rodrigues formula as
\begin{equation}
 P_l(x)=\frac{1}{2^l l!}\frac{d^l}{dx^l}(x^2-1)^l.
\end{equation}

In the representation of the Dirac equation eigenstates in Eq.~\eqref{eq:psi}, the radial functions $f_{n_\mathrm{r},\kappa_\mathrm{r}}(r)$ and $g_{n_\mathrm{r},\kappa_\mathrm{r}}(r)$ are given by
\begin{align}
 f_{n_\mathrm{r},\kappa_\mathrm{r}}(r) &=\frac{(2\lambda)^{3/2}}{\Gamma(2\gamma+1)}(2\lambda r)^{\gamma-1}e^{-\lambda r}\nonumber\\
 &\hspace{0.3cm}\times\sqrt{\frac{(C_1m_\mathrm{e}c^2+E_{n_\mathrm{r},\kappa_\mathrm{r}})\Gamma(2\gamma+n_\mathrm{r}+1)}{4C_1m_\mathrm{e}c^2\big(\frac{C_1Z\alpha_\mathrm{e} m_\mathrm{e}c^2}{C_2\lambda\hbar c}\big)\big(\frac{C_1Z\alpha_\mathrm{e} m_\mathrm{e}c^2}{C_2\lambda\hbar c}-\kappa\big)n_\mathrm{r}!}}\nonumber\\
 &\hspace{0.3cm}\times\Big[\Big(\frac{C_1Z\alpha_\mathrm{e} m_\mathrm{e}c^2}{C_2\lambda\hbar c}-\kappa\Big)\;F_{\hspace{-1em}1\hspace{0.7em}1}(-n_\mathrm{r};2\gamma+1;2\lambda r)\nonumber\\
 &\hspace{0.3cm}-n_\mathrm{r}\;\;F_{\hspace{-1em}1\hspace{0.7em}1}(1-n_\mathrm{r};2\gamma+1;2\lambda r)\Big],
 \label{eq:f}
\end{align}
\begin{align}
 g_{n_\mathrm{r},\kappa_\mathrm{r}}(r) &=\frac{-(2\lambda)^{3/2}}{\Gamma(2\gamma+1)}(2\lambda r)^{\gamma-1}e^{-\lambda r}\nonumber\\
 &\hspace{0.3cm}\times\sqrt{\frac{(C_1m_\mathrm{e}c^2-E_{n_\mathrm{r},\kappa_\mathrm{r}})\Gamma(2\gamma+n_\mathrm{r}+1)}{4C_1m_\mathrm{e}c^2\big(\frac{C_1Z\alpha_\mathrm{e} m_\mathrm{e}c^2}{C_2\lambda\hbar c}\big)\big(\frac{C_1Z\alpha_\mathrm{e} m_\mathrm{e}c^2}{C_2\lambda\hbar c}-\kappa\big)n_\mathrm{r}!}}\nonumber\\
 &\hspace{0.3cm}\times\Big[\Big(\frac{C_1Z\alpha_\mathrm{e} m_\mathrm{e}c^2}{C_2\lambda\hbar c}-\kappa\Big)\;F_{\hspace{-1em}1\hspace{0.7em}1}(-n_\mathrm{r};2\gamma+1;2\lambda r)\nonumber\\
 &\hspace{0.3cm}+n_\mathrm{r}\;\;F_{\hspace{-1em}1\hspace{0.7em}1}(1-n_\mathrm{r};2\gamma+1;2\lambda r)\Big].
 \label{eq:g}
\end{align}
Here $\Gamma(x)$ is the gamma function, and $\;\,F_{\hspace{-1em}1\hspace{0.7em}1}(a;b;x)$ is the Kummer confluent hypergeometric function. The auxiliary quantities $\lambda$ and $\gamma$ used in the definitions of the radial functions in Eqs.~\eqref{eq:f} and \eqref{eq:g} are given by
\begin{equation}
 \lambda=\frac{\sqrt{C_1^2m_\mathrm{e}^2c^4-E_{n_\mathrm{r},\kappa_\mathrm{r}}^2}}{C_2\hbar c},\hspace{0.5cm}
 \gamma=\sqrt{\kappa_\mathrm{r}^2-(Z\alpha_\mathrm{e})^2}.
\end{equation}

\vspace{0.1cm}\noindent\textbf{Acknowledgements}\\
This work has been funded by the Research Council of Finland under Contract No.~349971.

\vspace{0.1cm}\noindent\textbf{Author contributions}\\
M.P. performed the theoretical calculations and wrote the first draft of the manuscript. J.T. commented on the manuscript and participated in the interpretation of the results.

\vspace{0.1cm}\noindent\textbf{Competing interests}\\
The authors declare no competing interests.

\end{document}